\documentclass{iau}

\usepackage{amsmath}
\usepackage{multirow}

\def \jnl{}

\def\actaa{\jnl{AcA}}               
\def\aap{\jnl{A\&A}}                
\def\apj{\jnl{ApJ}}                 
\def\mnras{\jnl{MNRAS}}             
\def\apjl{\jnl{ApJL}}               
\def\nat{\jnl{Nature}}              

\begin{document}

\lefttitle{Bogumi{\l} Pilecki}
\righttitle{Cepheids with giant companions}

\jnlPage{1}{9}
\jnlDoiYr{2023}
\volno{376}
\pubYr{2024}
\doival{10.1017/S1743921323004258}

\aopheadtitle{Proceedings IAU Symposium}
\editors{Richard de Grijs, Patricia Whitelock and M\'arcio Catelan, eds.}

\title{Cepheids with giant companions: A new, abundant source of Cepheid astrophysics} 

\author{Bogumi{\l} Pilecki}
\affiliation{Centrum Astronomiczne im. Miko{\l}aja Kopernika, PAN, Bartycka 18, 00-716 Warsaw, Poland\\
  \email{pilecki@camk.edu.pl}}

\begin{abstract}
We present a progress report of our project aiming to increase the
number of known Cepheids in double-lined binary (SB2) systems from six
to 100 or more. This will allow us, among other goals, to
accurately measure masses for a large sample of Cepheids. Currently, only
six accurate Cepheid masses are available, which hinders our
understanding of their physical properties and renders the Cepheid
mass--luminosity relation poorly constrained. At the same time,
Cepheids are widely used for essential measurements (e.g.,
extragalactic distances, the Hubble constant). To examine Cepheid
period--luminosity relations, we selected as binary candidates
Cepheids that are too bright for their periods. To date, we have
confirmed 56 SB2 systems, including the detection of significant
orbital motions of the components for 32. We identified systems with
orbital periods up to five times shorter than the shortest reported
period to date, as well as systems with mass ratios significantly
different from unity (suggesting past merger events). Both features
are essential to understand how multiplicity affects the formation and
destruction of Cepheid progenitors and what effect this has on global
Cepheid properties. We also present eight new systems composed of two
Cepheids (only one such system was known before). Among confirmed SB2
Cepheids, there are also several wide-orbit systems. In the future,
these may facilitate independent accurate geometric distance
measurements to the Large and Small Magellanic Clouds.
\end{abstract}


\maketitle

\section{Introduction} \label{sec:intro}

Classical Cepheids are perhaps the most important objects in
astrophysics, given that they are crucial to gain improved physical
insights in various fields, including as regards stellar oscillations
and the evolution of intermediate-mass and massive stars. They wield
enormous influence on modern cosmology. Since the discovery of the
relationship between their pulsation period and their luminosity more
than a century ago (the Leavitt Law; \citealt{1912HarCi.173....1L}),
Cepheids have been used extensively to measure distances both within
and outside of our Galaxy. The recent local Hubble constant
determination accurate to 1.8\% \citep{Riess_2022_Hubble_1.4}, which
shows a significant discrepancy with the value inferred from {\sl
  Planck} data, depends sensitively on this period--luminosity (P--L)
relation.

Although theoretical studies of Cepheids are quite advanced (e.g.,
\citealt{Bono_1999_Cep_TheoryII,Bono_2005_ApJ_CEP_puls_model_IV,Valle_2009_AA_theo_pred_CC_puls}),
our empirical knowledge is still limited. Theory predicts masses of
Cepheids in the range of 3--11 M$_\odot$, but their measured masses
clump between 3.6 M$_\odot$ and 5 M$_\odot$, with only one higher but
uncertain value of 6 M$_\odot$
\citep{cep227mnras2013,allcep_pilecki_2018,Evans_2018_V350SGr_Mass,Gallene_2019AA_Galactic_Cepheids}. This
makes the Cepheid mass--luminosity relation very poorly constrained
\citep{Anderson_2016_Rotation}. Nevertheless, it is crucial for our
theoretical understanding of the P--L relation and basic stellar
physics regarding, e.g., convection, mass loss, and
rotation. Moreover, the blue loops predicted by current evolution
theory are too short to explain the existence of low-mass,
short-period Cepheids (see, e.g., Espinoza-Arancibia and Pilecki,
these proceedings).

Masses covering a wider range would be of the utmost importance to
resolve these issues, but we in practice can measure them only for
Cepheids in spectroscopic double-lined binaries (SB2), for which lines
of companions are easily detectable. Unfortunately, such systems are
very rare. Most binary Cepheids, and all found in the Milky Way, have
an early-type main-sequence companion (exhibiting few and broad lines)
which is typically 2--5 mag fainter in the $V$ band, making it
extremely hard to determine its velocity and thus the mass of the
Cepheid.

One good source for more binary Cepheids would be to look for
eclipses, but this solution has three serious limitations. First, such
systems are not numerous, because a very specific orbital orientiation
is needed for eclipses to occur. Second, according to current
statistics, only about half of eclipsing systems involving Cepheids
are double-lined, and third, the best places to look for them (the
Galactic disk, Magellanic Clouds) have already been examined, and we
do not expect many more such binaries to be found there any time
soon. Therefore, we decided to look for another possible source that
had not yet been considered and remained unexplored.

In the first paper of the series \citep{cepgiant1_2021} we described
such a source, highlighting its potential, and we presented the first
results for the Large Magellanic Cloud (LMC). Here, we present the
current stage of the project, providing updates based on new
observations and expansion of the sample to the Small Magellanic Cloud
(SMC) and the Milky Way (MW).

\section{The project}
\label{sec:project}

The basic idea of the project was to look for Cepheids in binary
systems, for which lines of both components are present in their
spectra. To meet these conditions, one has to find Cepheids
accompanied by stars of similar luminosity and preferentially of late
spectral types, i.e., subgiants or later stages of stellar
evolution. To identify such candidates, we considered three observable
features caused by such companions. Compared with single Cepheids (or
Cepheids with significantly fainter companions), for Cepheids with
giant companions we expect:

\begin{itemize}
\item their total observed brightness to increase significantly;
\item their photometric pulsation amplitude to decrease; and
\item their color to be either similar or redder.
\end{itemize}

\begin{figure}
    \begin{center}
        \includegraphics[width=\textwidth]{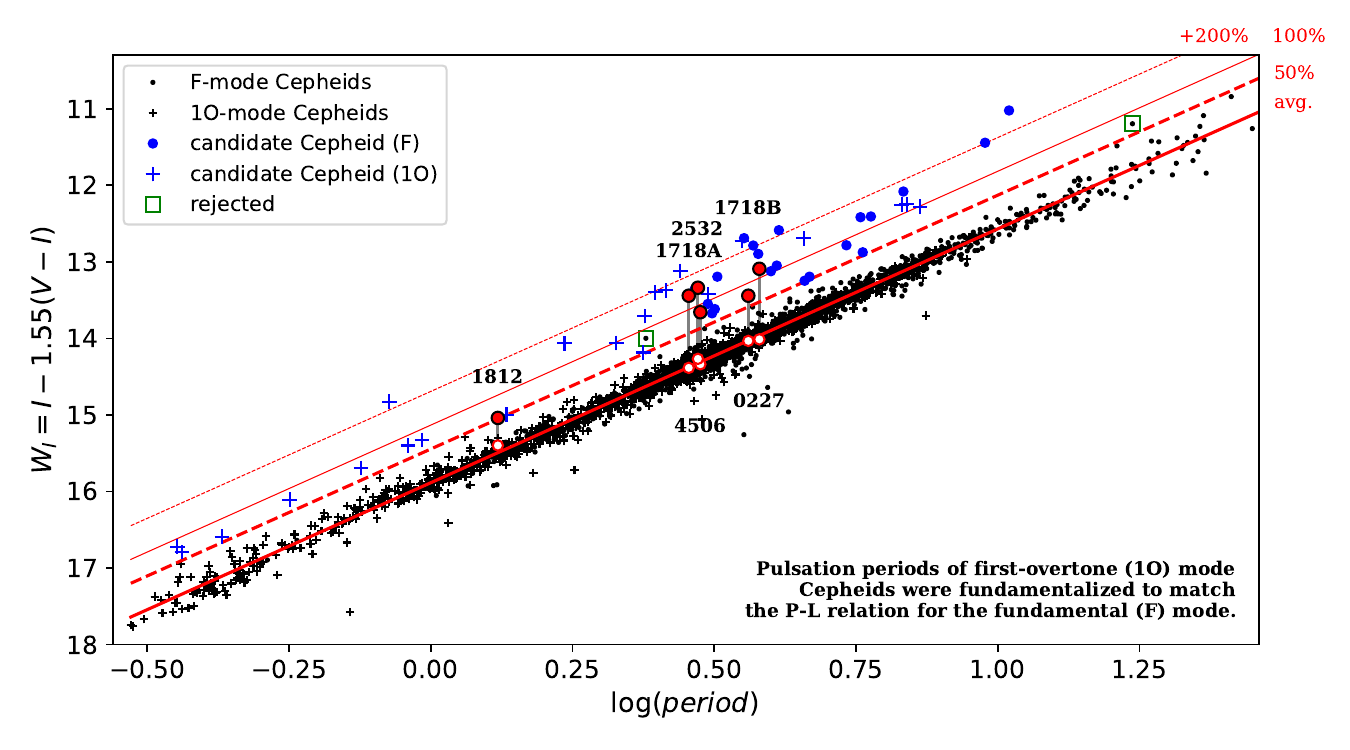} 
    \end{center}
\caption{P--L relation based on the reddening-free Wesenheit
  index. Filled red circles show known eclipsing binaries involving
  Cepheids, while empty circles mark their brightness without any
  companion light. Our candidates for binary Cepheids are $\ge$50\%
  brighter than an average Cepheid for a given period. Note that their
  distribution parallel to the P--L relation advocates strongly
  against blending by random stars.}
\label{fig:perlum}
\end{figure} 

Looking at the P--L diagram for LMC Cepheids
(Figure~\ref{fig:perlum}), we noticed that all previously confirmed
eclipsing giant--giant SB2 systems including Cepheids
\citep{allcep_pilecki_2018} lie significantly above the corresponding
P--L relation, being at least 50\% (0.44 mag) brighter than a typical
Cepheid for its period. At the same time, their amplitudes are about
half the typical ones, and their colors are typical or redder. From
the number of known eclipsing binary Cepheids we also estimated how
many similar binary Cepheids may exist but have not yet been detected
in photometric studies due to a lack of eclipses. Assuming random
inclinations, we expected on the order of 50 SB2 systems composed of a
Cepheid and a giant companion exhibiting a range of orbital periods
from 1 to $\sim$5 years, i.e., similar to the periods for eclipsing
Cepheids.

Indeed, when we investigated the P--L diagram for all Cepheids from
the OGLE-3 catalog \citep{Soszynski_2008_LMC_Cepheids}, we identified
41 additional Cepheids that lie 0.44 mag (4.7$\sigma$) above the P--L
relation, but for which eclipses were not detected
(Figure~\ref{fig:perlum}). Virtually all Cepheids selected this way
have low amplitudes and colors that are on average redder than typical
by 0.12 mag in the $(V-I)$ index. As stated above, these three
features together clearly indicate that the overbright Cepheids have
luminous, late-type giant companions. This, in turn, made them perfect
candidates for SB2 systems composed of giants, for which lines of both
components could be easily detected and their radial velocities (RVs)
measured.

The Cepheids in our sample are distributed along the full P--L
relation, with periods from 0.26 to 10.5 days. Among them, 20 pulsate
in the fundamental (F) mode and 21 in the first-overtone (1O)
mode. Three of the latter are actually double-mode Cepheids pulsating
in the second overtone (2O) as well. In Figure~\ref{fig:perlum} the
periods of 1O Cepheids ($P_{1O}$) were fundamentalized to match the
periods of F-mode Cepheids ($P_{F}$) using the prescription of
\citet{cepgiant1_2021}.

Meanwhile, the project has been expanded by inclusion of new
candidates in the LMC from the OGLE-4 catalog
\citep{Soszynski_2017AcA_OCVS_MC_Cep} and a dozen candidates in the
SMC \citep{Soszynski_2010_SMC_Cepheids}. In the samples selected as
described above we found all seven known double Cepheids (i.e.,
objects for which pulsation of two Cepheids was detected at the same
coordinates), which were all selected as candidates. This statistical
result suggests that they share the same properties as other
P--L-overbright Cepheids. This allows us to add to our SB2 candidate
list double Cepheids from the MW, where the P--L relation method of
detecting SB2 Cepheids cannot be applied directly (individual
distances to the objects are needed to do so).

\section{Data}
\label{sec:data}

Spectroscopic monitoring of our first SB2 candidates started in
October 2020, with new objects added subsequently. Observations were
performed using three of the world's best instruments mounted on
telescopes located at observatories in Chile. The brighest targets ($V
\leq 15.7$ mag) were observed with the High Accuracy Radial velocity
Planet Searcher (HARPS) instrument mounted on the 3.6 meter telescope
at the La Silla observatory. Most of the acquired spectra were
obtained with the Magellan Inamori Kyocera Echelle (MIKE) spectrograph
mounted on the 6.5 m Magellan Clay telescope at the Las Campanas
observatory. A significant fraction of data were also acquired with
the European Southern Observatory's UV--Visual Echelle Spectrograph
(UVES) on the 8.2 m Very Large Telescope (VLT) at Paranal
Observatory. HARPS and MIKE were used in visitor mode, so that the
observations were confined to the specific nights of the corresponding
runs. UVES observations were obtained in service mode and could thus
cover the orbital cycles more uniformly, while several-day-long HARPS
runs were useful for covering pulsation phases of bright, long-period
Cepheids.

\section{Results}
\label{sec:results}

\begin{figure}
    \begin{center}
        \includegraphics[width=0.6\textwidth]{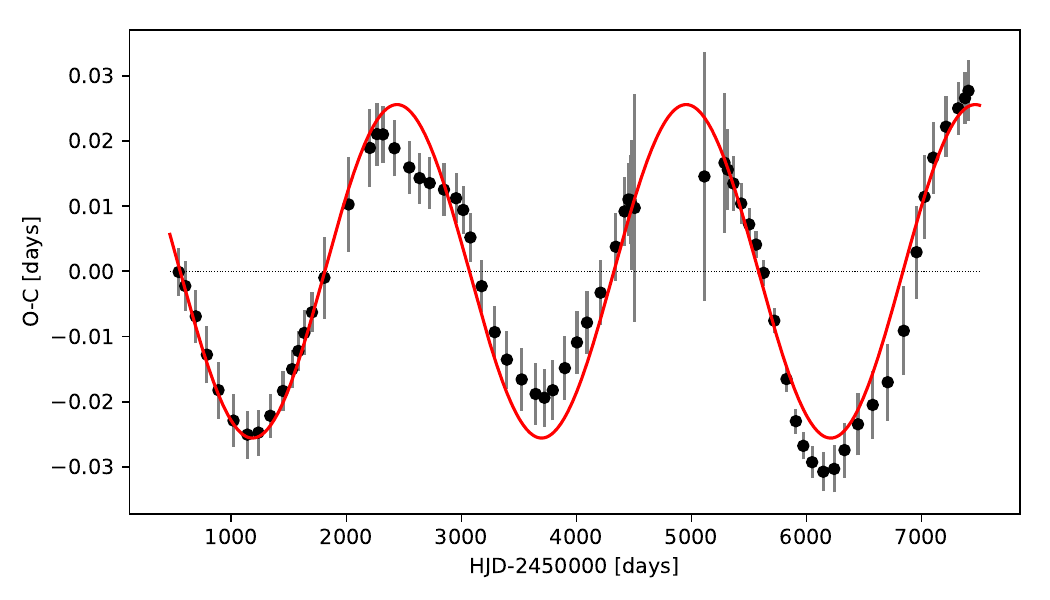}
    \end{center}
\caption{O--C diagram for one of the SB2 Cepheid candidates. A clear
  LTTE modulation due to binary motion is present. The red line is the
  best fit to the Cepheid's orbit.}
\label{fig:ltte0889}
\end{figure} 

In a preliminary study, we analyzed photometric data from the OGLE
catalogs for our initial sample of 41 Cepheids. The study of their
O--C diagrams led to detection of the light travel-time effect (LTTE)
due to binary motion for four candidates. One example, with an orbital
period of about 2500 days, is shown in Figure~\ref{fig:ltte0889}.

\subsection{Double-lined binary Cepheids}
\label{sec:sb2ceps}

\begin{figure}
    \begin{center}
        \includegraphics[width=0.56\textwidth]{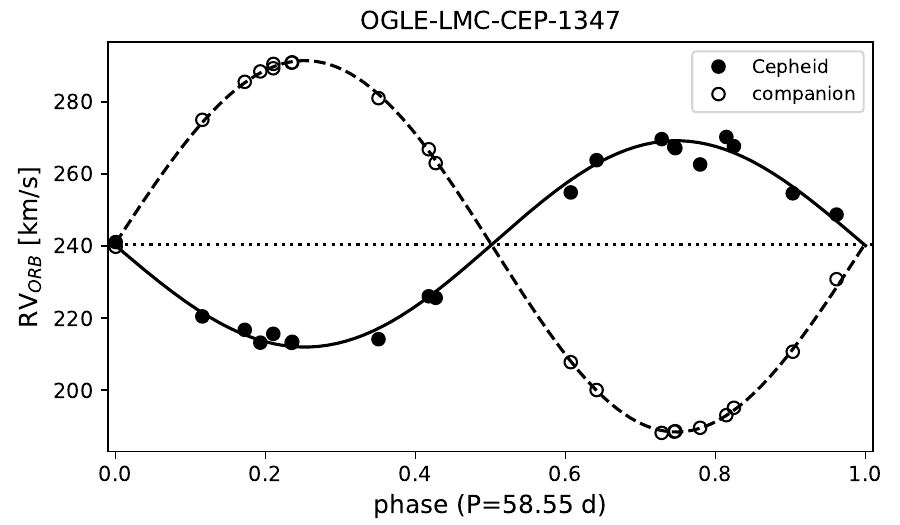}
        \includegraphics[width=0.43\textwidth]{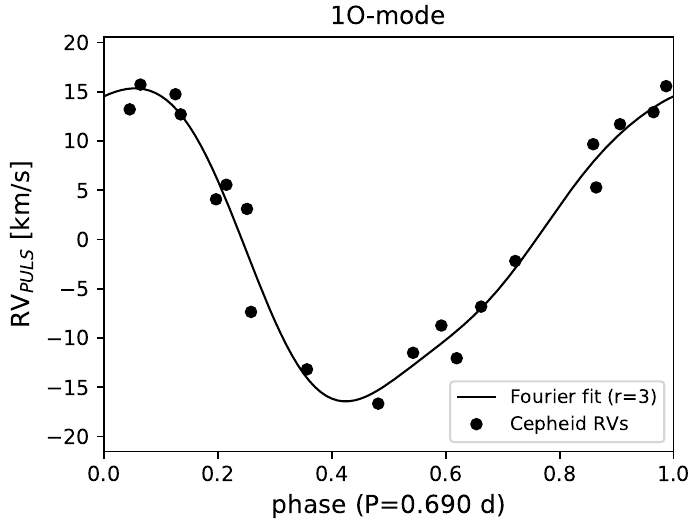}
    \end{center}
\caption{Updated RV curves for OGLE-LMC-CEP-1347. The scatter in the
  Cepheid RVs is due to unaccounted-for 2O-mode pulsations.}
\label{fig:cep1347}
\end{figure}

During the pilot program, we found 16 of 18 ($\sim$90\%) of our
brightest candidate Cepheids to be components of SB2 systems, showing
that the majority of these overbright pulsators are normal Cepheids
with luminous giant companions. Additional observations of fainter
candidates led to the discovery of a Cepheid in a binary system with
an orbital period of only 59 days, five times shorter than any
measured before for a firmly confirmed binary Cepheid. For a more
detailed description of the system, please refer to
\citet{cep1347_ApJL_2022}. Updated RV curves for this object are shown
in Figure~\ref{fig:cep1347}.

\begin{figure}
    \begin{center}
        \includegraphics[width=0.6\textwidth]{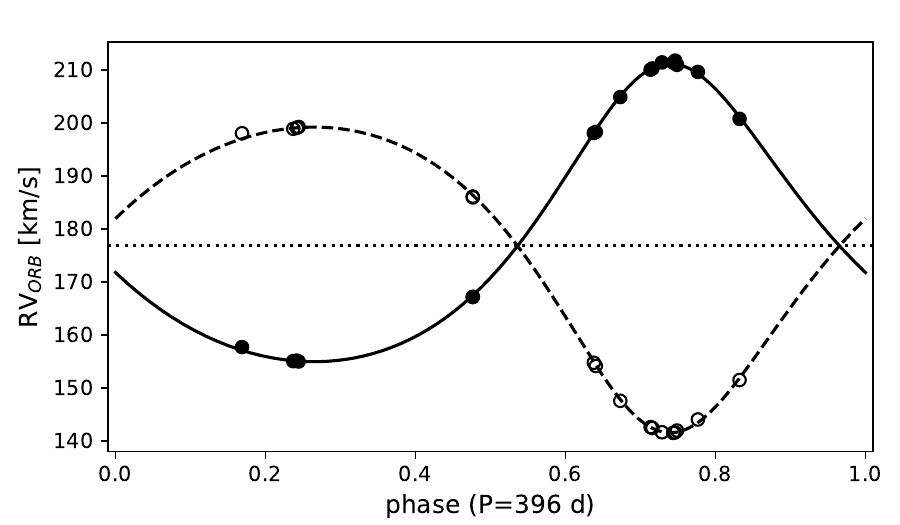}
        \includegraphics[width=0.6\textwidth]{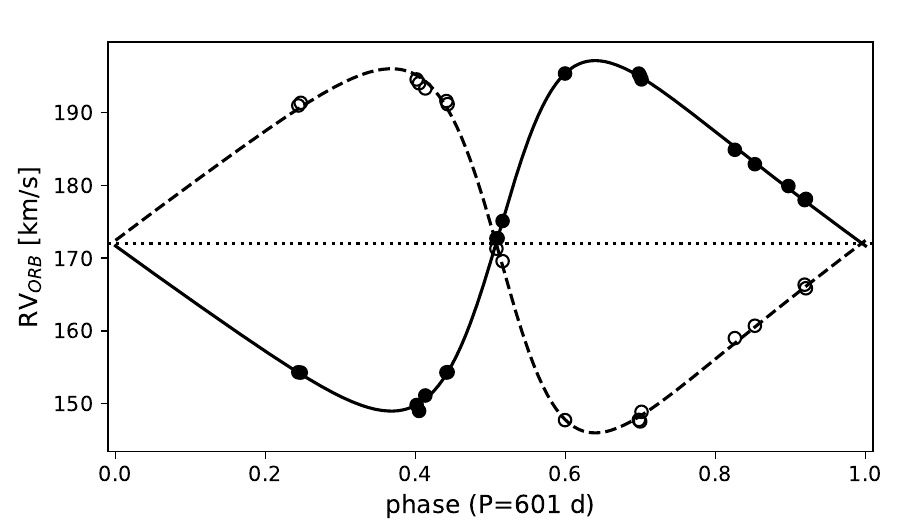}
    \end{center}
\caption{Preliminary orbits for two example double-lined binary
  Cepheids with best-covered orbital cycles. More data are still
  needed to cover the pulsation curves and improve the precision of
  the derived orbital parameters but no major change is expected to
  the solution.}
\label{fig:sb2ceps}
\end{figure} 

The observations have continued and, to date, 56 candidate Cepheids in
the LMC, SMC, and MW have been confirmed as members of SB2
systems. For 32 of them, the anticorrelated orbital motions of the
Cepheid and its companion were detected, which represents the ultimate
proof that they are gravitationally bound. For 23, preliminary orbital
solutions could be obtained, although in general the data coverage is
not yet satisfactory due to long orbital periods and the complexity of
the variabilities involved. For a few cases we managed to cover the
orbital cycles well, and the solutions are not expected to change
significantly (see Figure~\ref{fig:sb2ceps}). Moreover, eight of the
confirmed systems are composed of two Cepheids (see the next
subsection).

\begin{figure}
    \begin{center}
        \includegraphics[width=0.6\textwidth]{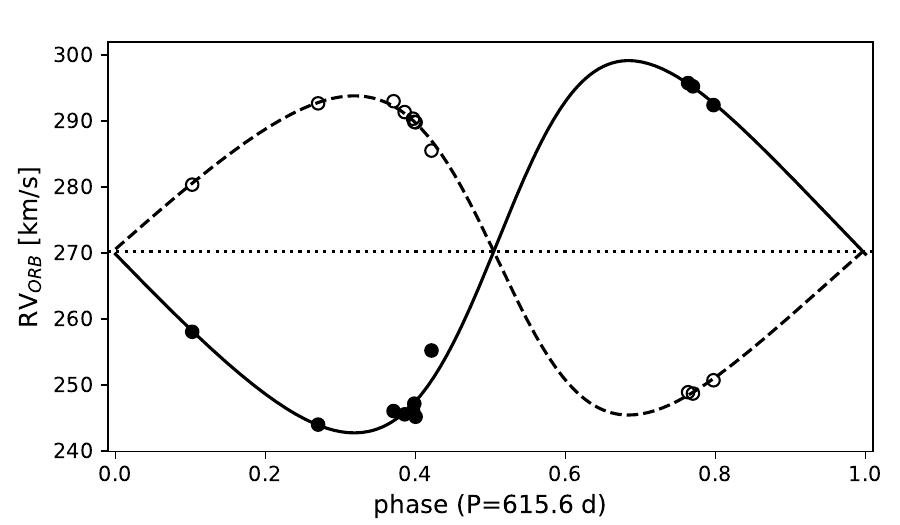}
    \end{center}
\caption{Example of a binary Cepheid with a preliminary mass ratio
  that is significantly different from unity.}
\label{fig:qnot1}
\end{figure}

For systems with preliminary orbital solutions we calculated the mass
ratios of the components and the minimum Cepheid masses, $M_{\rm
  Cep}\sin^3(i)$. The minimum masses obtained for several systems are
close to the masses derived for eclipsing Cepheids, showing that the
inclinations of these systems are high and that they are good targets
for accurate and precise mass measurements. Mass ratios obtained for
some of the systems are different from unity. As $q=M_2/M_1 \sim 1$ is
expected for components at similar evolutionary stages and the same
age, which means that they could have passed through binary
interactions in the past, either by means of a merger event or via
mass transfer. One example of such a system is shown in
Figure~\ref{fig:qnot1}.

\begin{figure}
    \begin{center}
        \includegraphics[width=0.6\textwidth]{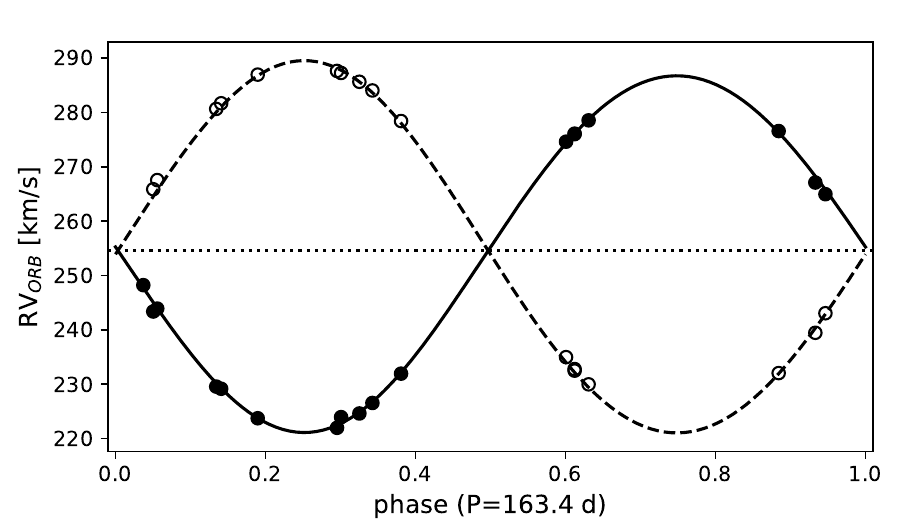}
    \end{center}
\caption{Another binary Cepheid with an orbital period below the
  200-day limit.}
\label{fig:shortper}
\end{figure}

Among the binary Cepheids with orbital solutions, a new system was found
with an orbital period below 200 days, which is the lower limit obtained by \citet{Neilson2015_cep_binsys_porb} for binary
Cepheids that passed through their evolution on the RGB. As systems with
periods below this limit would be destroyed at that evolutionary
stage, their components are probably younger, and the Cepheids are
likely on their first crossing through the instability strip (in the
Hertzsprung gap). Our newly discovered system has a period of only 163 days,
which is the second shortest period ever measured for a binary Cepheid
(Figure~\ref{fig:shortper}).

\begin{figure}
    \begin{center}
        \includegraphics[width=0.6\textwidth]{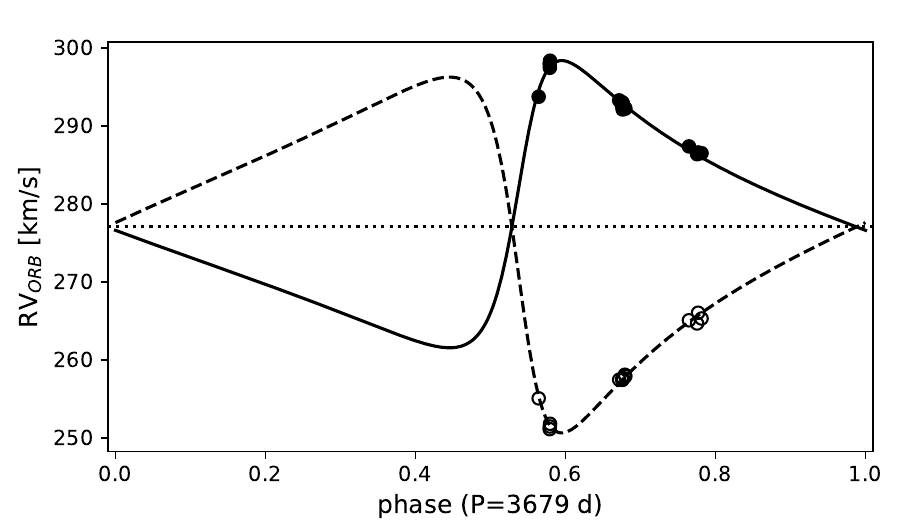}
    \end{center}
\caption{Example of a wide-orbit binary Cepheid with a long orbital
  period. The current solution suggests that its period may reach 10
  years or even longer.}
\label{fig:longper}
\end{figure}

The longest orbital period measured so far for an extragalactic binary system is $P=1550$ days ($\sim$4 yr; \citealt{cep9009apj2015}).
Meanwhile, preliminary solutions for some of our new systems suggest orbital periods
longer than 4 years, even up to about 10 years or more (see
Figure~\ref{fig:longper}).
These binary systems, for which we can obtain spectroscopic orbits,
are currently the best candidates for direct geometric
distance measurements using interferometric methods, offering the best possible accuracy.
And although at the moment we do not expect high precision for
extragalactic measurements, with upcoming improvements to
instrumentation, such systems may eventually become our best and most
direct tools for distance measurements to nearby galaxies, surpassing
previously used methods (e.g., \citealt{Pietrzyn_2019_LMC_1perc}). 

\subsection{Binary double Cepheids}
\label{sec:bindceps}

\begin{figure}
    \begin{center}
        \includegraphics[width=0.6\textwidth]{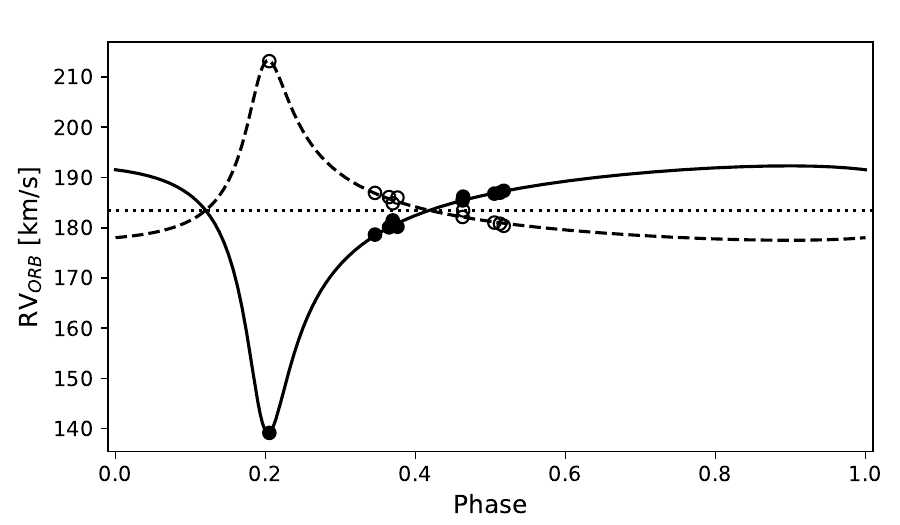}
        \includegraphics[width=0.6\textwidth]{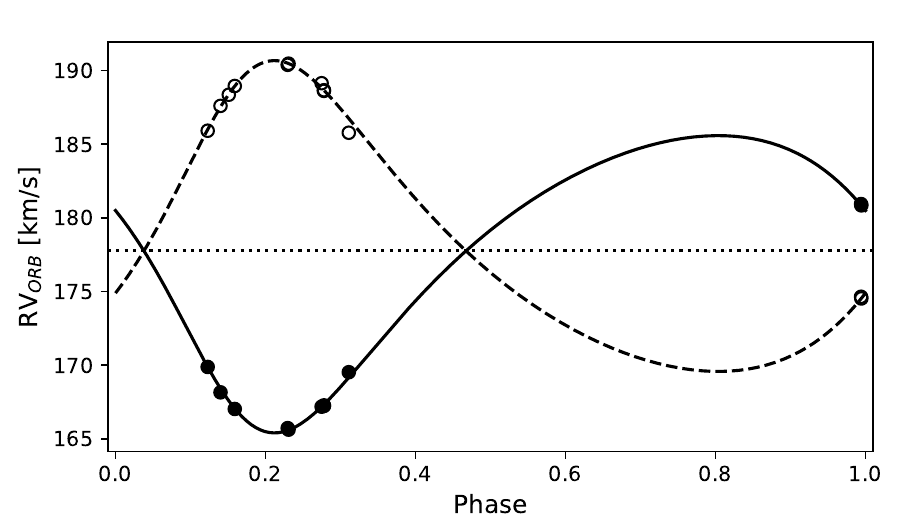}
    \end{center}
\caption{Preliminary orbits for two example binary double Cepheids
  show a clear change of the components' orbital velocities. However,
  as the orbital cycle has not yet been fully covered, the final,
  exact solutions (including the orbital periods) may change
  significantly.}
\label{fig:bindceps}
\end{figure}

An interesting subsample of double-lined binary Cepheids form those
binary systems composed of two Cepheid components. As part of our
project we observed nine candidates for such systems. Binary double
(BIND) Cepheids are unique laboratories for pulsation and evolution
studies. However, only one system of this kind was known before. To
date, we have confirmed orbital motions for eight additional systems
of a similar nature. Preliminary orbits for two are shown in
Figure~\ref{fig:bindceps}. Altough the orbital solutions may change
once a full orbital cycle has been covered, the anti-correlated
orbital motion of both components is already clear, proving binarity
of these double Cepheids.

Interestingly, some preliminary solutions suggest mass ratios
significantly different from unity, while for others the pulsation
periods are very different (after fundamentalization of the 1O
periods, for comparison). Both features may indicate past binary
interaction events, which would affect the physical properties of the
same-age components. Another interesting feature of the new BIND
Cepheids is their long orbital periods. For most, periods longer than
five years are expected according to the current models (based on only
partially covered cycles). Although long periods make the observations
more challenging, in the future these systems will also offer an
opportunity to measure direct distances to their host galaxies, as described in the previous section.
More recent results for BIND Cepheids can be found in \citet{bindceps_AA_2024}.

\section{Conclusions}
\label{sec:conclusions}

Analysis of spectroscopic data we have collected so far clearly shows
that the Cepheids selected as described in Section~\ref{sec:project}
are indeed components of double-lined binary systems. This has
important implications for the interpretation of period--luminosity
relations and for our general knowledge of Cepheids and their
evolution. We know now that overbright Cepheids, which are often
rejected as P--L relation outliers, are in the great majority Cepheids
with red, luminous giant companions. These conclusions can also likely
be extended to other pulsating stars with well-defined P--L relations,
e.g., Type II Cepheids or RR Lyrae stars.

With 56 confirmed candidates, we have already ten-fold increased the
number of Cepheids in spectroscopic double-lined binaries. This
includes over 90\% of all candidates in the LMC. In
Figure~\ref{fig:perlum} one can see that many of them have periods
longer than those of known SB2 Cepheids, reaching up to 10.5 days.
Extrapolation of the number of confirmed cases to the SMC suggests
that about 50 new SMC Cepheids in SB2 systems can be expected once all
SMC objects have been studied. This would increase the total number of
known SB2 Cepheids by a further 80\% (i.e., 20-fold compared with the
state before we commenced the project).

Eventually, our study will yield firm mass estimates for a large
sample of Cepheids, including long-period, high-mass Cepheids for
which the lack of data is the most severe and for those from the
uncertain low-mass end. Similar evolutionary stages for both
components imply mass ratios close to unity. Any significant deviation
from unity may indicate a probable past merger event. This is the only
way merged Cepheids can be unambiguously detected (as compared with,
e.g., dubious chemical composition peculiarities) and
characterized. From the simulations of
\citet{Sana_2012_BinInteractions} we can estimate that up to 30\% of
Cepheids may be merger products. In our study we will be able to
estimate this fraction observationally and compare it with
simulations.

Geometric methods provide the most accurate and direct distances to
astronomical objects. One such method is based on the determination of
both astrometric (angular) and spectroscopic (absolute) orbits of
binary systems, which combined give directly the distance (see, e.g.,
\citealt{Gallene_2019AA_eclipsing_binaries}). Unfortunately, at
increasingly large distances, it becomes commensurately harder to
measure orbital angular sizes, as one needs systems of increasingly
higher luminosity and larger component separations. Using traditional
methods, such systems are extremely hard to identify in other
galaxies, as one needs decades of monitoring to find just a
few. However, our method of detection of binary Cepheids is
independent of the orbital size, and several systems with very wide
orbits have already been identified. They are currently our best
candidates for direct geometric distance determinations to the LMC and
SMC, which can eventually lead to the ultimate calibration of the
first rung of the cosmic distance ladder.

\acknowledgements{The research leading to these results received
  funding from the Polish National Science Center (grant SONATA BIS
  2020/38/E/ST9/00486). This work is based on observations collected
  at the Las Campanas Observatory and the European Southern
  Observatory, ESO (ESO programs: 106.21GB, 108.229Z, 108.22BS,
  110.2434, and 110.2436). We thank the Carnegie Institution and the
  CNTAC for the allocation of observing time for this project. This
  research has made use of NASA's Astrophysics Data System Service.}


\end{document}